\documentclass{article}
\usepackage{graphicx}
\usepackage{cite}
\usepackage{amssymb,amsmath,latexsym,enumerate}
\usepackage{color}

\begin{document}

\Large

\bigskip

\centerline{\textbf{Lax pair and first integrals for two}}
\centerline{\textbf{of nonlinear coupled oscillators}}

\bigskip

\centerline{Nikolay A. Kudryashov\footnote{\emph{Corresponding author:} nakudr@gmail.com}}

\bigskip

\normalsize

\centerline{National Research Nuclear University MEPhI (Moscow Engineering Physics}
\centerline{ Institute), 31 Kashirskoe Shosse, 115409 Moscow, Russian Federation}

\bigskip

\hrulefill

\textbf{Abstract}

The system of two nonlinear coupled oscillators is studied. As partial case this system of equation is reduced to the Duffing oscillator which has many applications for describing physical processes. It is well known that the inverse scattering transform is one of the most powerful methods for solving the Cauchy problems of partial differential equations. To solve the Cauchy problem for nonlinear differential equations we can use the Lax pair corresponding to this equation. The Lax pair for ordinary differential or systems or for system ordinary differential equations allows us to find the first integrals, which also allow us to solve the question of integrability for differential equations. In this report we present the Lax pair for the system of coupled oscillators. Using the Lax pair we get two first integrals for the system of equations. The considered system of equations can be also reduced to the fourth-order ordinary differential equation and the Lax pair can be used for the ordinary differential equation of fourth order. Some special cases of the system of equations are considered.

\bigskip


\hrulefill

\section{Introduction}{\label{Int}}

It is known that Gardner, Green. Kruskal and Miura first opened the inverse scattering transform \cite{Gardner, Lamb, Drazin} for solving the Cauchy problem of the Korteweg-de Vries equation \cite{Korteweg}. Using a linear system of equations of the above-mentioned authors Peter Lax  in 1968 introduced a new concept \cite{Lax, Kudr_19A, Kudr_19I} now called the Lax pair, which allows to solve the Cauchy problem by means of the inverse scattering transform for a certain class of nonlinear evolution equations.

Five years later, in 1973 four young graduates from the Potsdam University Mark Ablowitz, David Kaup, Alain Newell and Harvi Segur suggested to look for nonlinear evolution equations for which the Cauchy problems can be solved by the inverse scattering transform taking into account the operator equation.

Using the power dependencies of the matrix elements on the spectral parameter and on the function and their derivatives, from the operator equation for the AKNS scheme
e dependencies of the matrix elements and the evolutionary equations are  sequentially looked for which the Cauchy problem is solved by the inverse scattering transform.
Certainly we assume that if a nonlinear differential equation passes the Painlev\'e test, then the necessary condition for the integrability of an ordinary nonlinear differential equation is satisfied \cite{Ablow_73, Ablow_74, Kudr_18A, Kudr_19J, Ablow, Gromak}.

The disadvantage of using the Painlev\'e test for nonlinear differential equations is that despite the useful information contained in the Fuchs indices and in the expansions of the General solutions to the Laurent series found in the Painlev\'e test, as a result we obtain neither a general solution of the differential equation nor its first integrals \cite{Kudr_98, Kudr_14, Kudr_19}.

The aim of this paper is to look for nonlinear integrable ordinary differential equations and the first integrals using the modified AKNS scheme for nonlinear ordinary differentia equations.

The rest of this work is organized as follows. In Section \ref{Lax pair} we discuss the Lax pair associated with the system of nonlinear ordinary differential eqquations. In Section \ref{Second} using the Lax pair we find the system of two nonlinear ordinary differential equations of the second order. We also find the first integrals for this system of equations and discuss the partial cases.

\section{The pair for the system of nonlinear coupled equations}{\label{Lax pair}}

Let us consider the following system of nonlinear differential equations
\begin{equation}\begin{gathered}
\label{System_1}
a_1\,q_{tt}+b_1\,q_t+c_1\,p\,q^2+d_1\,q=0
\end{gathered}\end{equation}
\begin{equation}\begin{gathered}
\label{System_2}
a_2\,p_{tt}+b_2\,p_t+c_2\,q\,p^2+d_2\,p=0,
\end{gathered}\end{equation}
where $\rm p(t)$ and $\rm q(t)$ are unknown functions and $t$ is independent variable, $a_1$$a_2$$b_1$,$b_2$, $c_1$, $c_2$, $d_1$ and $d_2$ are parameters of mathematical models.

Let us look for the Lax pairs for the system of equations in the form
\begin{equation}\begin{gathered}
\label{Lax}
\rm \textbf{A}\,\psi=\lambda\,\psi,\\
\rm \psi_t=\textbf{B}\,\psi,
\end{gathered}\end{equation}
where $\rm \psi$, $\rm \textbf{A}$ and $\rm \textbf{B}$ are matrices in the form \cite{Ablow_73,Ablow_74,Ablow}
\begin{equation}\begin{gathered}
\label{M1}
\rm {\psi}=$$ \rm \begin{pmatrix}
\rm \,\psi_1 & \, \\
\\
\rm \, \psi_2 & \, \end{pmatrix}$$, \qquad
\rm \textbf{A}=$$ \rm \begin{pmatrix}
\rm a_{11} & & \,\, \rm a_{12}\\
\\
\rm a_{21} & & \,\, \rm a_{22}\end{pmatrix}$$, \qquad
\rm $$ \rm \textbf{B}=\begin{pmatrix}
\rm -i\,\lambda & & \rm \,\,q(t)\\
\\
\rm p(t) & & \,\,\rm i\,\lambda \end{pmatrix}$$
\end{gathered}\end{equation}
We look for the Lax pair for traveling wave reduction of the KdV hierarchy takin into account the equation
\begin{equation}\begin{gathered}\label{M}
\rm \frac{d{\textbf{A}}}{dt}=\textbf{B}\,\textbf{A}-\textbf{A}\,\textbf{B}.
\end{gathered}\end{equation}
Note that equation \eqref{M} similar to the Lax pair for the KdV hierarchy if we write this one using the traveling wave solutions.

From equation \eqref{M} we have four ordinary differential equation for matrix elements  $\rm a_{11}$, $\rm a_{12}$, $\rm a_{21}$ and $\rm a_{22}$  in the form
\begin{equation}\begin{gathered}\label{ODE_F_1}
\rm \frac{d {a}_{11}}{dt}=w\,a_{21}\,-\,w\,a_{12},
\end{gathered}\end{equation}
\begin{equation}\begin{gathered}\label{ODE_F_2}
\rm \frac{d {a}_{12}}{dt}\,=-2\,i\,\lambda\,a_{12}\,+w\,a_{22}\,-w\,\,a_{11},
\end{gathered}\end{equation}
\begin{equation}\begin{gathered}\label{ODE_F_3}
\rm \frac{d {a}_{21}}{dt}\,=\,w\,a_{11}-w\,a_{22}\,+2\,i\,\lambda\,a_{21},
\end{gathered}\end{equation}
\begin{equation}\begin{gathered}\label{ODE_F_4}
\rm \frac{d {a}_{22}}{dt}\,= \,w\,a_{12}-w\,a_{21}.
\end{gathered}\end{equation}
Adding equations \eqref{ODE_F_1} and \eqref{ODE_F_4} we have
\begin{equation}\begin{gathered}\label{ODE_F_5}
\rm \frac{d }{dt}\,(a_{11} + {a}_{22})\,= \,0.
\end{gathered}\end{equation}
From the last equality follows that we get
\begin{equation}\begin{gathered}\label{ODE_F_5}
\rm a_{11} =- {a}_{22}.
\end{gathered}\end{equation}
Taking into account equations \eqref{ODE_F_2} and \eqref{ODE_F_3} we obtain
\begin{equation}\begin{gathered}\label{ODE_F_6}
\rm \frac{d}{dt}(a_{12}+a_{21})=2\,i\,\lambda\,(a_{21}-a_{12}).
\end{gathered}\end{equation}
Let us look for the dependence of elements $\rm a_{11}$, $\rm a_{12}$, $\rm a_{21}$ and $\rm a_{22}$  in the form
\begin{equation}\begin{gathered}\label{Elements}
\rm {a}_{11}=\sum_{k=0}^{n}\,a_k(w,\,w_z,\ldots)\,\lambda^{n-k}, \quad {a}_{12}=\sum_{k=0}^{n-1}\,b_k(w,\,w_z,\ldots)\,\lambda^{\,n-1-k},
\\
\rm {a}_{21}=\sum_{k=0}^{n-1}\,c_k\,(w,\,w_z,\ldots)\,\lambda^{\,n-1-k}.\qquad \rm {a}_{22}=-a_{11}
\end{gathered}\end{equation}
The matrix elements $\rm a_{11}$, $\rm a_{12}$, $\rm a_{21}$ and $\rm a_{22}$ of the matrix $\rm \textbf{A}$ can be used for finding the first integrals of the system of equations.
It is known that if the matrix $\rm \textbf{A}$ satisfies equation \eqref{M} then the first integrals corresponding to the original equation can be obtained by means of calculating of  traces $\rm tr \textbf{A}^k$.

We can use the consequence of this proposition.
If the matrix elements $\rm a_{11}=-a_{22}$ then $\rm tr \textbf{A}^2=-2\, det {\textbf{A}}$. Let us
note that in case $\rm a_{11}=-a_{22}$ we have the following equality
\begin{equation}\begin{gathered}\label{Pr_1}
\rm tr \textbf{A}^2=a_{11}^2+2\,a_{12}a_{22}+a_{22}^2=(a_{11}+a_{22})^2+\\
\\
\rm +2\,a_{12}\,a_{21}-2\,a_{11}\,a_{22}=0+2(a_{12}\,a_{21}-a_{11}\,a_{22}) =-2\, det {\textbf{A}}.
\end{gathered}\end{equation}
So, to look for the first integrals of the system of equations \eqref{System_1} and \eqref{System_2}we  have to calculate the determinant of matrix $\rm \textbf{A}$.

\section{Two nonlinear coupled oscillators and their first integrals}{\label{Second}}

Let us assume  in \eqref{Elements} $\rm n=2$. In this case we have
\begin{equation}\begin{gathered}
\label{Elements_2}
\rm a_{11}(t)=a_2(t)+a_1(t)\,\lambda+a_0(t)\,\lambda^2,\qquad a_{22}=-a_{11},\\
\\
\rm a_{12}(t)=b_1(t)+b_0(t)\,\lambda,\qquad a_{21}(t)=c_1(t)+c_0(t)\,\lambda.
\end{gathered}\end{equation}
Substituting \eqref{Elements_2} into equations \eqref{ODE_F_1}, \eqref{ODE_F_2}, \eqref{ODE_F_3} and \eqref{ODE_F_4} we have after calculations the following values of the matrix elements
\begin{equation}\begin{gathered}
\label{Elements_a11}
\rm a_{11}=\,-\alpha\,p(t)\,q(t)-\,C_0-i\,\beta\,\,\lambda-2\,\alpha\,\lambda^2, \qquad a_{22}=-a_{11},
\end{gathered}\end{equation}
\begin{equation}\begin{gathered}
\label{Elements_a12}
\rm a_{12}=\alpha\,q_t+\,\beta\,q\,-2\,i\,\alpha\,q\,\lambda,\qquad  a_{21}=-\,\alpha\,p_t+\,\beta\,p\,-2\.i\,\alpha\,p\,\lambda.
\end{gathered}\end{equation}
We also have the system of equations in the form
\begin{equation}\begin{gathered}
\label{Eq_2}
\rm \alpha\,q_{tt}+\,\beta\,q_t-2\,\alpha\,p\,q^2\,-2\,C_0\,q=0
\end{gathered}\end{equation}
and
\begin{equation}\begin{gathered}
\label{Eq_1}
\rm {\alpha}\,p_{tt}-\,\beta\,p_t-2\,\alpha\,\,p^2\,q-2\,C_0\,p=0,
\end{gathered}\end{equation}
where $\rm p(t)$ and $\rm q(t)$ are unknown functions and $\rm \alpha$, $\rm\beta$ and $\rm C_0$ are parameters of the system of equations.

To look for the first integrals for the system of equations from the Lax pair we have to calculate the determinant of matrix $\rm \textbf{A}$.
Determinant of matrix $\rm \textbf{A}$ takes the form
\begin{equation}\begin{gathered}\label{Pr_2}
\rm det{\textbf{A}}={\alpha^2}\,p_t\,q_t-{\alpha^2}\,p^2\,q^2-2\,C_0\,\alpha\,p\,q-\,C_0^2-\beta^2\,p\,q+\\
\\
\rm +\,\alpha\,\beta\,(q\,p_t-p\,q_t)+2\,i\,\left(\alpha\,\beta\,p\,q-\beta\,C_0+{\alpha^2}p\,q_t-{\alpha^2}q\,p_t\right)\lambda+\\
\\
+(\beta^2-4\,\alpha\,C_0)\,\lambda^2-4\,i\,\alpha\,\beta\,\lambda^3-4\,\alpha^2\,\lambda^4.
\end{gathered}\end{equation}
From expression \eqref{Pr_2} we obtain two first integral for the system of equations \eqref{Eq_1} and \eqref{Eq_2} in the form
\begin{equation}\begin{gathered}\label{F_2B}
\rm I_1=\alpha\,\beta\,p\,q-\beta\,C_0+{\alpha^2}p\,q_t-{\alpha^2}q\,p_t
\end{gathered}\end{equation}
and
\begin{equation}\begin{gathered}\label{F_2A}
\rm I_2={\alpha^2}p_t q_t-{\alpha^2} p^2 q^2-2 C_0 \alpha\,p q-\beta^2 p q+\alpha \beta\, (q p_t-p q_t)- C_0^2.
\end{gathered}\end{equation}
Now let us consider the partial cases of the system of equations \eqref{Eq_1} and \eqref{Eq_2} with obtained lax pair.

Assuming $\rm \alpha=1$, $\rm \beta=0$ and $\rm p=q$ we have the well-known second-order  nonlinear differential equation
\begin{equation}\begin{gathered}\label{Eq_3}
\rm q_{tt}-2\,q^3-2\,C_0\,q=0.
\end{gathered}\end{equation}
As this takes place integral \eqref{F_2B} is generated and integral \eqref{F_2A} is transformed to the well-known integral for equation \eqref{Eq_3} in the form
\begin{equation}\begin{gathered}\label{Eq_3AB}
\rm I_2^{(1)}=q_{t}^2-\,q^4-\,C_0\,q^2=C_2.
\end{gathered}\end{equation}
The general solution of equation \eqref{Eq_3AB} is expressed by mens of the Jacobi elliptic function.

Equation \eqref{Eq_3AB} can be written in the form
\begin{equation}\begin{gathered}
\label{Integ_3AB}
\rm q_t^2\,=(q-\alpha)\,(q-\beta)\,(q-\gamma)\,(q-\delta).
\end{gathered}\end{equation}
where $\alpha$, $\rm \beta$, $\rm \gamma$ and $\rm \delta$ ($\rm \alpha \geq \beta \geq \gamma \geq \delta$ ) are real roots of the algebraic equation
\begin{equation}\begin{gathered}
\label{Roots_1}
\rm q^4+\frac12\,C_0\,q^2+C_2+0.
\end{gathered}\end{equation}
Equation \eqref{Integ_3AB} can be transformed to the following form
\begin{equation}\begin{gathered}
\label{Integ_3A}
\rm v_t^2=(1-v^2)(1-k^2\,v^2), \\
\\
\rm v^2=\frac{(\beta-\delta)(q-\alpha)}{(\alpha-\delta)(y-\beta)},\quad k^2=\frac{(\beta-\gamma)(\alpha-\delta)}{(\alpha-\gamma)(\beta-\delta)}
\end{gathered}\end{equation}
The general solution of \eqref{Integ_3AB} is expressed via
the elliptic function in the form \cite{Davis, Kudr_19C, Kudr_19D, Kudr_19E, Kudr_19G}
\begin{equation}\begin{gathered}
\label{Sol_Kdv_3}
\rm v(t)= sn\left(\chi t,k \right),\qquad \chi^2=\frac14\,(\beta-\delta)\,(\alpha-\gamma)
\end{gathered}\end{equation}
where $\rm sn\left(\chi t,k \right)$ is the elliptic sine.

The general solution of equation \eqref{Sol_Kdv_3} takes the form
\begin{equation}\begin{gathered}
\label{Solitary_1}
\rm y(t)=\frac{\beta(\alpha-\delta)\,sn^2\left(\chi t,k \right)-\alpha(\beta-\delta)}{(\alpha-\delta)\,sn^2\left(\chi t,k \right)-\beta+\delta}.
\end{gathered}\end{equation}

From equation \eqref{Eq_2} we get
\begin{equation}\begin{gathered}\label{Eq_4}
\rm p=\frac{q_{tt}}{2\,q^2}+\frac{\beta\,q_t}{2\,\alpha\,q^2}-\frac{C_0}{\alpha\,q}.
\end{gathered}\end{equation}
Substituting $\rm p$ from \eqref{Eq_4} into equation \eqref{Eq_3} we have the fourth-order differential equation in the form
\begin{equation}\begin{gathered}\label{Eq_5}
\rm \alpha^2\,\left(q^2\,q_{tttt}-4\,q\,q_z\,q_{ttt}-3\,q\,q_{tt}^2+6\,q_t^2\,q_{tt}\right)+\\
\\
\rm +\alpha\,\left(6\beta\,q_t^3-6\,\beta q\,q_t\,q_{tt}+4\,C_0\,q^2\,q_{tt}-4\,C_0\,q\,q_t^2\right) \\
\\
\rm +{\beta^2}\,q\,q_t^2-{\beta^2}\,q^2\,q_{tt}=0.
\end{gathered}\end{equation}
From \eqref{F_2A} and \eqref{F_2B} we have the first integrals for \eqref{Eq_5} in the form
\begin{equation}\begin{gathered}\label{F_3B}
\rm I_1=\alpha^2\,\left(q\,q_{ttt}-3\,q_t\,q_{tt}\right)+\alpha\,\left(4 C_0 q q_t-3\,\beta q_t^2\right)+4\,\beta\,C_0\,q^2-\beta^2\,q\,q_t
\end{gathered}\end{equation}
and
\begin{equation}\begin{gathered}\label{F_3A}
\rm I_2=\alpha^3\left(2\,qq_tq_{ttt}-4\,q_t^2\,q_{tt}-q\,q_{tt}^2\right)+4\,\beta^2\,C_0q^3-2\,\beta^3q^2\,q_t+\\
\\
\rm +\alpha^2 \left(2 \beta q^2 q_{ttt}-6\beta q q_t q_{tt}-4 \beta q_t^3+4 C_0qq_t^2\right)+\alpha \beta \left( 8 C_0 q^2 q_t-7 \beta q q_t^2 \right).
\end{gathered}\end{equation}
Assuming $\rm \beta=0$ and $\rm \alpha=1$ we obtain from \eqref{Eq_2} and \eqref{Eq_1} the system of equations for description in the form
\begin{equation}\begin{gathered}
\label{Eq_11}
\rm \,p_{tt}-2\,p^2\,q-2\,C_0\,p=0
\end{gathered}\end{equation}
and
\begin{equation}\begin{gathered}
\label{Eq_22}
\rm q_{tt}-2\,q^2\,p-2\,C_0\,q=0
\end{gathered}\end{equation}
with Hamiltonian that follows from the first integral \eqref{F_2A} in the form
\begin{equation}\begin{gathered}
\label{Hamil}
\rm H= p_t\,q_t-\,q^2\,p^2-2\,C_0\,q\,p=0.
\end{gathered}\end{equation}
The first integrals $\rm I_1$ and $\rm I_2$ at $\rm lpha=1$ and $\rm \beta$ take the form
\begin{equation}\begin{gathered}\label{FF_2B}
\rm I_1=-p\,q_t-q\,p_t
\end{gathered}\end{equation}
and
\begin{equation}\begin{gathered}\label{FF_2A}
\rm I_2={}p_t q_t-{} p^2 q^2-2 C_0 p q - C_0^2.
\end{gathered}\end{equation}
Assuming
\begin{equation}\begin{gathered}\label{FF_2A}
\rm  q_1=q,\qquad p_1=p_t,\qquad q_2=p, \qquad p_2=q_t
\end{gathered}\end{equation}
we obtain that system equations \eqref{Eq_11} and \eqref{Eq_22} are the Hamilton system of equations
\begin{equation}\begin{gathered}
\label{Ham_1}
\rm \dot{q}_i=\frac{\partial H}{\partial p_i},\qquad \dot{p}_i=-\frac{\partial H}{\partial q_i}, \qquad (i\,=\,1,\,\,2).
\end{gathered}\end{equation}
We also obtain that the first integrals \eqref{F_2B} and \eqref{F_2A} satisfy to the involution, As this take we have
\begin{equation}\begin{gathered}
\label{Involution}
\rm \{I_1,\,\,I_2\}=0
\end{gathered}\end{equation}
At $\rm \beta\neq0$ the system equations \eqref{Eq_2} and \eqref{Eq_1} is the Hamilton system too. hamiltonian for this system of equation can be found form the first integrals \eqref{F_2B} and \eqref{F_2A} using the same variables $\rm q_i$ and $\rm p_i$, where  $(i\,=\,1,\,\,2)$.

We have at $\alpha=1$ and $\beta=0$ the following integrable  differential equation of fourth order
\begin{equation}\begin{gathered}\label{Eq_6}
\rm q^2\,q_{tttt}-4\,q\,q_t\,q_{ttt}-3\,q\,q_{tt}^2+6\,q_t^2\,q_{tt}+4\,C_0\,q^2\,q_{tt}-4\,C_0\,q\,q_t^2=0
\end{gathered}\end{equation}
with two first integrals in the form
\begin{equation}\begin{gathered}\label{F_4B}
\rm I_1=q\,q_{ttt}-3\,q_t\,q_{tt}+4\,C_0\,q\,q_t
\end{gathered}\end{equation}
and
\begin{equation}\begin{gathered}\label{F_4A}
\rm I_2=2\,q\,q_t\,q_{ttt}-4\,q_{tt}\,q_t^2-q\,q_{tt}^2 +4\,C_0\,q\,q_t^2.
\end{gathered}\end{equation}
Equation \eqref{F_4B} can be interated with respect to $\rm z$. it takes the form
\begin{equation}\begin{gathered}\label{F_4BB}
\rm q\,q_{tt}-2\,q_t^2 +2\,C_0\,q^2=I_1\,t+I_3
\end{gathered}\end{equation}
Taking into account the new variable $\rm q=-\frac{1}{V} $ we have from \eqref{F_4BB} the second order differential equation in the form
\begin{equation}\begin{gathered}\label{New_Eq2}
\rm V_{tt}-2\,C_0\,V+\left(I_3+I_1\,t\right)\,V^3=0,
\end{gathered}\end{equation}
where $\rm I_1$ and $\rm I_3$ are arbitrary constants.

At $\rm I_1=0$ we obtain after integration the equation for the eliiptic function Jacobi in the form
\begin{equation}\begin{gathered}\label{F_4CC}
\rm V_{t}^2-2\,C_0\,V^2+\frac12\,I_3\,V^4=C_4,
\end{gathered}\end{equation}
where $C_4$ is arbitrary constant.

Let us use the new variable $\rm q(t)=-\frac{1}{V(t)}$ again in equation \eqref{F_4A}. We have
\begin{equation}\begin{gathered}\label{New_Eq2A}
\rm 2\,V_t^2\,V_{tt}-V\,V_{t}\,V_{ttt}+\frac12\,V\,V_{tt}^2
-4\,C_0\,V\,V_z^2-\,I_2\,V^6=0
\end{gathered}\end{equation}
Substituting $\rm V_{tt}$ from \eqref{New_Eq2} we get the first-order nonlinear equation in the form
\begin{equation}\begin{gathered}\label{New_Eq_3}
\rm \left(I_1\,t+I_3\right)V_t^2 +I_1\,V\,V_t-2\,C_0^2-\left( 2\,I_1\,C_0\,t+I_3\,C_0\,\right)\,V^2-\\
\\
\rm -\frac12I_2\,V^3+\left(\frac12\,I_3^2+I_1\,I_3\,t+\frac12\,I_1^2\,t^2\right)\,V^4=0.
\end{gathered}\end{equation}
From \eqref{New_Eq_3} at $\rm I_1=0$ we obtain the equation
\begin{equation}\begin{gathered}\label{New_Eq_4}
\rm \rm I_3\,V_t^2 +2\,C_0^2-2\,I_3\,C_0\,V^2 -\frac12I_2\,V^3+\frac12\,I_3^2\,V^4=0..
\end{gathered}\end{equation}
The general solution of equation \eqref{New_Eq_4} is expressed via the Jacobi elliptic function.

\section{Conclusion}{\label{Concl}}

In this report we have considered the system of two nonlinear differential equations. We have found the Lax pair for this system. Using this one we have obtained the first integrals for the system of equations.

\section*{Acknowledgements}

The reported study was funded by RFBR according to the research Project No. 18-29-10025.


\begin{thebibliography}{99}


\bibitem{Gardner} \textit{C. S. Gardner, J. M. Green, M. D. Kruskal, R. M. Miura}, Method for solving Korteweg-de Vries equation. Phys. Rev. Lett. {\bf{19}} (1967) 1095-1097.

\bibitem{Lamb} \textit{G. L. Lamb}, 1980, {\it Elements of Soliton Theory}, (New York, John Wiley and Sons).

\bibitem{Drazin} \textit{P.G. Drazin, R.S. Johnson}, 2002, {\it Solitons: an Introduction}, (Cambridge University Press).


\bibitem{Korteweg} \textit{D.J. Korteweg, G. de Vries}, On the change of form of long waves advancing in a rectangular canal and a new tipe of long stationary waves, Phi. Mag.  {\bf{39}} (1895) 422-423.

\bibitem{Lax} \textit{P.D. Lax}, Integrals of nonlinear equation of evolution and solitary waves, Pure Appl. Math. {\bf{21}} (1968) 467-490.

\bibitem{Kudr_19A} \textit{N. A. Kudryashov}, Lax pair and first integrals of the traveling wave reduction for the KdV hierarchy, Applied Mathematics
and Computation {\bf 350} (2019) 323--330.

\bibitem{Kudr_19I} \textit{N. A. Kudryashov}, Traveling wave reduction for the modified KdV hierarchy: Lax pair and first integrals, Communications in
Nonlinear Science and Numerical Simulat {\bf 73} (2019) 472--480.

\bibitem{Ablow_73} \textit{M.J. Ablowitz, D.J. Kaup, A.C. Newell, H. Segur}, Nonlinear evolution equations of physical significance, Phys. Rev. Lett. {\bf{31}} (1973) 125-127.

\bibitem{Ablow_74} \textit{M.J. Ablowitz, D.J. Kaup, A.C. Newell, H. Segur}, The inverse scattering transfor -- Fourieranalysis for nonlinear problem , Stud.Appl.Math. {\bf{53}} (1974) 249--315.

\bibitem{Kudr_18A} \textit{N. A. Kudryashov}, Exact Solutions and Integrability of the Duffing - Van der Pol Equation, Regular and Chaotic Dynamics {\bf 23 (4)} (2018) 471--479.

\bibitem{Kudr_19J} \textit{N. A. Kudryashov}, Exact solutions of the equation for surface waves in a convecting fluid, Applied Mathematics and Computation {\bf 344} (2019) 97--106.

\bibitem{Gromak} \textit{V.I. Gromak}, {Painlev\'e Differential Equations in the Complex Plane}, Walter de Gruyter, Berlin, New York, 2002, 300.

\bibitem{Ablow} \textit{M.J. Ablowitz, P.A. Clarkson}, {Solitons, Nonlinear Evolution Equations and Inverse Scattering}, Cambridge University Press, 1991, 516.

\bibitem{Kudr_98} \textit{N.A. Kudryashov}, On new transcendents defined by nonlinear ordinary differential equations, J.Phys.A: Math.Gen, {\bf31} (1998) L129--L137.

\bibitem{Kudr_14} \textit{N.A. Kudryashov}, Higher Painlev\'e Transcensents as Special Solutions of Some Nonlinear Integrable Hierarchies, Regular and Chaotic Dynamics, {\bf 19(1)}, (2014) 48--63.

\bibitem{Kudr_19} \textit{N.A. Kudryashov}, Nonlinear differential equations associated with the first Painlev\'e herarchy, Applied Mathematics Letters {\bf 91} (2019) .

\bibitem{Davis} \textit{H.T. Davis}, Introduction to Nonlinear Differential and Integral Equations, Dover, New York,1962.

\bibitem{Kudr_19C} \textit{N.A. Kudryashov}, First integrals and solutions of the traveling wave reduction for the Triki-Biswas equation, Optik, 185 (2019) 275--281.

\bibitem{Kudr_19D} \textit{N.A. Kudryashov}, First integrals and general solution of the traveling wave reduction for the Schrodinger equation
with anti-cubic nonlinearity, Optik, 185 (2019) 665--671.

\bibitem{Kudr_19E} \textit{N.A. Kudryashov},	General solution of the traveling wave reduction for the Kundu-Mukherjee-Naskar model, Optik, 186 (2019) 22--27.

\bibitem{Kudr_19G} \textit{N.A. Kudryashov}, A generalized model for description of propagation pulses in optical fiber, Optik, 189 (2019) 42--52.






\end{thebibliography}
\end{document}